%% file: main.tex
\documentclass[sigconf]{acmart}
\usepackage{fancybox}

\AtBeginDocument{%
  \providecommand\BibTeX{{%
    \normalfont B\kern-0.5em{\scshape i\kern-0.25em b}\kern-0.8em\TeX}}}

\setcopyright{acmcopyright}
\copyrightyear{2018}
\acmYear{2018}
\acmDOI{XXXXXXX.XXXXXXX}

\acmConference[Conference acronym 'XX]{Make sure to enter the correct
  conference title from your rights confirmation emai}{June 03--05,
  2018}{Woodstock, NY}
\acmPrice{15.00}
\acmISBN{978-1-4503-XXXX-X/18/06}

\newcommand{\mx}[1]{\textcolor{black}{\textbf{#1}}}
\usepackage{graphicx}
\usepackage{tabularx}
\begin{document}

\title{Form 10-K Itemization}

\author{Yanci Zhang}
\authornote{Disclaimer: Work was done prior to joining Amazon.}
\affiliation{%
  \institution{Wharton Research Data Services}
  \city{Philadelphia}
  \country{USA}
}
\email{yanci@wharton.upenn.edu}

\author{Mengjia Xia}
\affiliation{%
  \institution{Wharton Research Data Services}
  \city{Philadelphia}
  \country{USA}
}
\email{xiax@wharton.upenn.edu}

\author{Mingyang Li}
\affiliation{%
  \institution{University of Pennsylvania}
  \city{Philadelphia}
  \country{USA}
}
\email{myli@alumni.upenn.edu}

\author{Haitao Mao}
\affiliation{%
 \institution{Michigan State University}
 \state{Michigan}
 \country{USA}}
\email{haitaoma@msu.edu}

\author{Yutong Lu}
\affiliation{%
  \institution{University of Oxford}
  \city{Oxford}
  \country{UK}
  }
\email{yutong.lu@mansfield.ox.ac.uk}

\author{Yupeng Lan}
\affiliation{%
  \institution{The University of Hong Kong}
  \city{Hong Kong}
  \country{China}
}
\email{stevelan@connect.hku.hk}

\author{Jinlin Ye}
\affiliation{%
  \institution{Penn Law}
  \institution{The Wharton School}
  \institution{University of Pennsylvania}
  \city{Philadelphia}
  \country{USA}}
\email{jinlinye@upenn.edu}

\author{Rui Dai}
\authornote{Corresponding author.}
\affiliation{%
  \institution{Wharton Research Data Services}
  \institution{The Wharton School}
  \institution{University of Pennsylvania}
  \city{Philadelphia}
  \country{USA}}
\email{rdai@wharton.upenn.edu}

\renewcommand{\shortauthors}{Yanci et al.}

\begin{abstract}
  \input{body/abstract.tex}
\end{abstract}





\maketitle

\input{body/10K-structure.tex}
\input{body/intro.tex}

\input{body/background.tex}

\input{body/challenge.tex}

\input{body/methodology.tex}

\input{body/result.tex}

\begin{figure}[t]
    \centering
    \includegraphics[width=\linewidth]{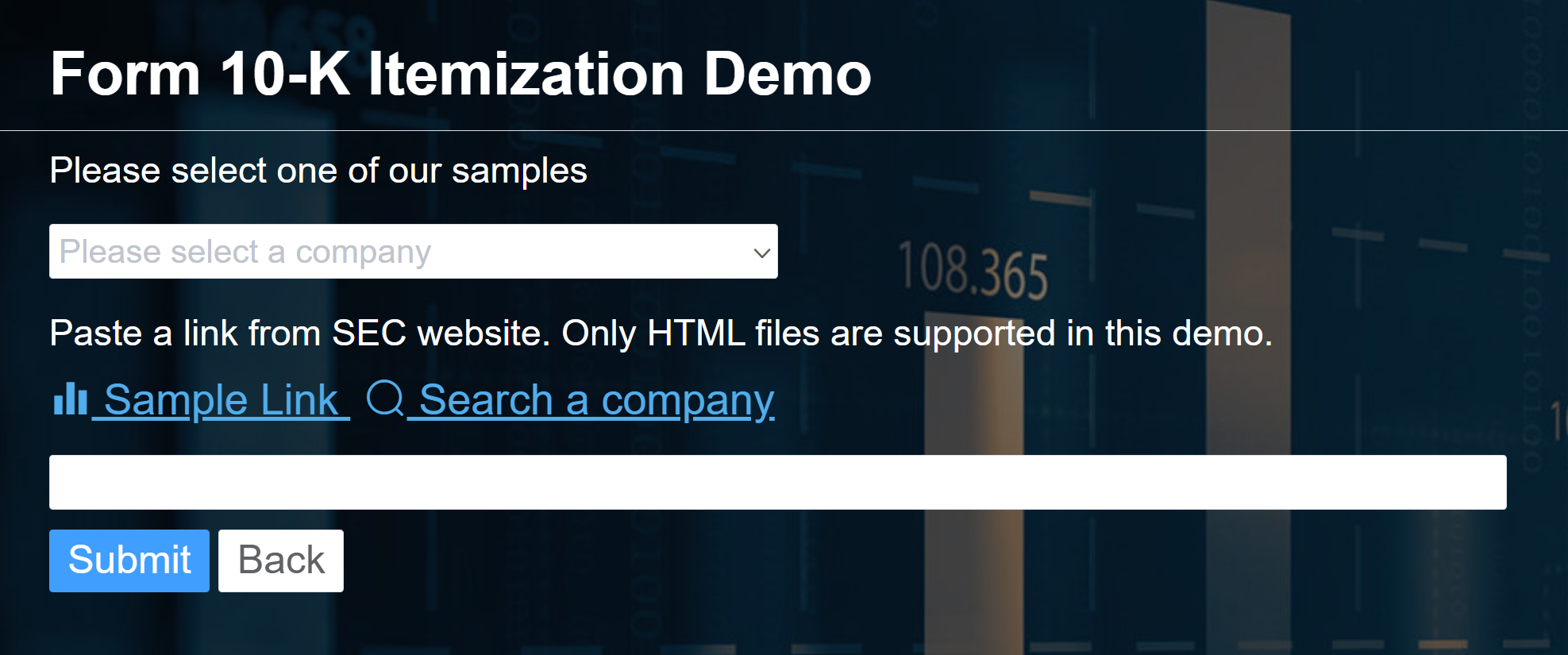}
    \vskip 1em
    \caption{Interface of Form 10-K Itemization system. Users can choose a sample filing or provide an SEC link to perform the itemization.}
    \label{fig:demo}
\end{figure}

\begin{figure}[t]
    \centering
    \includegraphics[width=\linewidth]{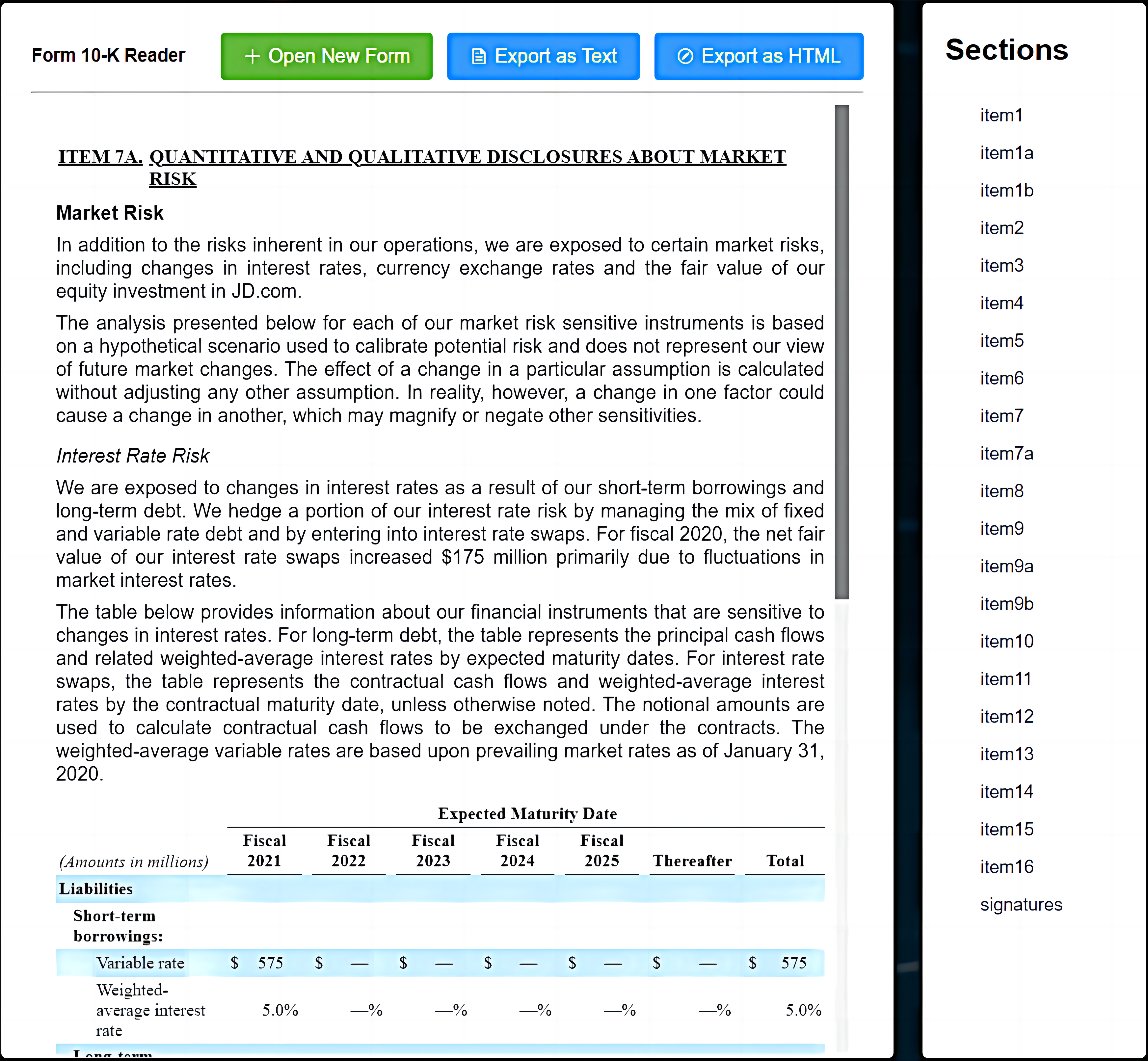}
    \vskip 1em
    \caption{Interface of Form 10-K Reader. Retrieved items are listed on the right and the content of each item is presented on the left.}
    \label{fig:demo:reader}
\end{figure}

\input{body/demo.tex}
\input{body/conclusion.tex}

\bibliographystyle{ACM-Reference-Format}
\bibliography{body/ref.bib}

\end{document}

%% file: body/abstract.tex
Form 10-K report is a financial report disclosing the annual financial state of a public company. It is an important evidence to conduct financial analysis, i.e., asset pricing, corporate finance. 
Practitioners and researchers are constantly designing algorithms to better conduct analysis on information in the Form 10-K report. The vast majority of previous works focus on quantitative data. With recent advancement on natural language processing (NLP), textual data in financial filing attracts more attention. However, to incorporate textual data for analyzing, Form 10-K  Itemization is a necessary pre-process step. It aims to segment the whole document into several Item sections, where each Item section focuses on a specific financial aspect of the company. With the segmented Item sections, NLP techniques can directly apply on those Item sections related to downstream tasks. In this paper, we develop a Form 10-K Itemization system which can automatically segment all the Item sections in 10-K documents. The system is both effective and efficient. It reaches a retrieval rate of 93\%. The system demo website can be found \href{http://review10-k.ddns.net}{here}.

%% file: body/10K-structure.tex
\begin{table}[t]
\caption{The SEC guidance \cite{tenk-guideline} on Form 10-K reports, including 4 parts and 22 items. Each item focuses on specific financial aspect of the company }
\label{tab:10-K}
\begin{tabular}{l}
\hline \hline
\textbf{PART I}                                                                                                                                                 \\ \hline
Item 1. Business.                                                                                                                                               \\
Item 1A. Risk Factors.                                                                                                                                          \\
Item 1B. Unresolved Staff Comments.                                                                                                                             \\
Item 2. Properties.                                                                                                                                             \\
Item 3. Legal Proceedings.                                                                                                                                      \\
Item 4. Mine Safety Disclosures.                                                                                                                                \\ \hline
\textbf{PART II}                                                                                                                                                \\ \hline
\begin{tabular}[c]{@{}l@{}}Item 5. Market for Registrant’s Common Equity, Related\\ Stockholder Matters and Issuer Purchases of Equity Securities.\end{tabular} \\
Item 6. {[}Reserved{]}                                                                                                                                          \\
\begin{tabular}[c]{@{}l@{}}Item 7. Management’s Discussion and Analysis of Financial\\ Condition and Results of Operations. \end{tabular}
                                                                \\
\begin{tabular}[c]{@{}l@{}}Item 7A. Quantitative and Qualitative Disclosures About Market\\ Risk.   \end{tabular}
                                                                                           \\
Item 8. Financial Statements and Supplementary Data.                                                                                                            \\
\begin{tabular}[c]{@{}l@{}}Item 9. Changes in and Disagreements With Accountants on\\ Accounting and Financial Disclosure.   \end{tabular}
                                                                 \\
Item 9A. Controls and Procedures.                                                                                                                               \\
Item 9B. Other Information.                                                                                                                                     \\
\begin{tabular}[c]{@{}l@{}}Item 9C. Disclosure Regarding Foreign Jurisdictions that Prevent\\ Inspections.  \end{tabular}
                                                                                 \\ \hline
\textbf{PART III}                                                                                                                                               \\ \hline
Item 10. Directors, Executive Officers and Corporate Governance.                                                                                                \\
Item 11. Executive Compensation.                                                                                                                                \\
\begin{tabular}[c]{@{}l@{}}Item 12. Security Ownership of Certain Beneficial Owners and\\ Management and Related Stockholder Matters.  \end{tabular}
                                                        \\
\begin{tabular}[c]{@{}l@{}}Item 13. Certain Relationships and Related Transactions, and\\ Director Independence.   \end{tabular}
                                                                            \\
Item 14. Principal Accountant Fees and Services.                                                                                                                \\ \hline
\textbf{PART IV}                                                                                                                                                \\ \hline
Item 15. Exhibit and Financial Statement Schedules.                                                                                                             \\
Item 16. Form 10–K Summary.   \\ \hline

\textbf{Signature} \\ \hline \hline                                                                                                                                 
\end{tabular}
\end{table}

%% file: body/intro.tex
\section{Introduction} \label{sec:intro}
The Form 10-K report is one of the most important documents in financial domains. It is an annual report providing a comprehensive summarization of the company’s financial state throughout the year. Following the guidance of the Securities and Exchange Commission (SEC), it should contain 4 parts and 22 Items as shown in Table~\ref{tab:10-K}. Each Item contains corresponding quantitative and textual data with different purposes. Quantitative data, which mainly lie in the tables, are easily recognized and well-studied in many financial applications \cite{fama1993common, fama2015five, gu2020empirical}. However, the largely ignored textual data, which occupies a significant part of documents, provide the opportunity for more comprehensive analysis. Nowadays, financial reports analysis is still labor-intensive which heavily relies on human experts. Therefore, how to automatically analyze with advanced natural language processing (NLP) has received increasing attention in recent years \cite{huang2011multilabel, li2013measure, yang2018corporate, mushtaq2022financial}.
   
Form 10-K Itemization is a task to segment the whole Form 10-K report into several Item sections shown in Table~\ref{tab:10-K}, where each Item section focuses on a specific financial aspect of the company. It is an essential prior step before incorporating natural language processing (NLP) techniques into financial report analysis. Its importance can be attributed to the following domain-specific reasons:

\begin{itemize}
    \item The predefined semantic structure of financial reports provides valuable instruction on analysis. However, it is largely ignored in the general NLP techniques. Financial reports are not unstructured text sequences but rather contain a predefined structure where each item section has its specific financial meaning. This structure can serve as important prior knowledge for financial analysis. However, it can be difficult for NLP techniques to extract useful information from the reports directly.
    \item The Form 10-K reports are typically long documents, while specific downstream analysis only relies on one or a few Item sections rather than utilizing the entire document. It is more suitable to apply NLP techniques on those specific Item sections while dropping other uninformative Item sections. 
\end{itemize}

By breaking down the form 10-K report into individual Item segmentations according to the guidance shown in Table \ref{tab:10-K}, we obtain structured smaller pieces of textual data with the specific financial topic. It sheds light on the following analysis: (1) chronological comparison of financial reports, which enables analysis of textual narratives throughout the years to identify trends and patterns. (2) a more granular and detailed analysis of financial information enhanced by specific-designed NLP techniques.

Form 10-K Itemization can be challenging for the following difficulties. The number of Form 10-K reports is large, with variations in format and semantics. It can be attributed to the flexibility in writing financial reports as follows (1) Flexibility on Item title: The company may not strictly write Item titles the same as the SEC suggested in Table \ref{tab:10-K}. The companies have flexibility in their presentation of different financial reports, making it difficult to identify Item titles; (2) Flexibility on which items to report: The company may not write every Item section if it is irrelevant to the financial state of the company. In other words, some Item sections cannot be found in particular reports, making it harder to identify the structure of the report. Some companies may also choose to use their own structure and style \cite{ge2021, mcd2020}. Another crucial challenge is that the number of labeled segmentation of Form 10-K reports is too small due to the label-intensive labeling procedure. It greatly complicates directly utilizing Deep Neural Networks for supervised learning. Deep models are most likely overfitting with unsatisfied performance. A more detailed discussion of challenges can be found in Section \ref{sec:challenge}.

A 10-K Itemization system is then proposed to itemize the Form 10-K reports. The itemization system is able to identify different sections of financial reports and convert them into a structured, computer-readable format\footnote{Demo tool URL: \url{http://review10-k.ddns.net}.}. The contents of the financial reports are stored as key-value pairs. Each output has a key \{Filing Serial Number\}\#\{Part Number\}\#\{Item Number\}, and value as content. 
Our system is a one-stop solution for financial filing itemization. It has the following superior features: 

\begin{enumerate}
    \item Full coverage: rather than extracting only one specific Item section \cite{loughran_when_2011, dyer_evolution_2017, liutracking}, the whole document, including all 22 Items in 10-K, is covered. 
    \item High accuracy: the itemization system has stable performance across companies and years with the help of expert knowledge and Deep model assisted automation. The retrieval rate on the whole document can reach 9xxx\% on the 10-K documents, while existing works on single item reports 75\% retrieval rate on Item 1 \cite{loughran_when_2011}, 27\% retrieval rate on item \mx{7} \cite{dyer_evolution_2017}. 
    \item Expert in the loop: Our system does not require a labor-intensive labeling process but a small expert labeled set. The system includes the capability for human-in-the-loop intervention with expert knowledge. It enables the automatic identification and highlighting of potential issues in the data, allowing for additional attention and review by human experts.
    \item Easy Extensibility: this proposed itemization system is compatible to perform well on other types of financial reports, i.e., Form 10-Q, though the discussion of this paper only focuses on evaluating 10-K.
    \item High availability: Our system deployment is designed to handle 10,000 requests per second from customers simultaneously, enabling large-scale, high availability analysis to occur at any time.
\end{enumerate}

The organization of the following sections is as follows: Section~\ref{sec:background} and Section~\ref{sec:relatedWork} introduce background and related work. Section~\ref{sec:challenge} emphasizes the key challenges in building our system. Section~\ref{sec:method} details the methodology used in the process pipeline. Section~\ref{sec:result} discusses the data set and provides detailed information on pipeline performance. Section~\ref{sec:demo} presents our demo interface and the design of our high-availability systems. Section~\ref{sec:conclusion} concludes our work and discusses future directions.

%% file: body/background.tex
\section{Background} \label{sec:background}

Financial reports provide comprehensive disclosure of current corporate governance and financial conditions of companies issuing securities in the United States. Under the Securities and Exchange Act, these security issuers are required to file periodic financial reports, including quarterly reports (Form 10-Q), audited annual reports (Form 10-K), etc., with the Securities and Exchange Commission (SEC). With such informativeness and reliability, these fillings are the primary source for practitioners and researchers to conduct financial analysis and research. 

SEC, as a regulatory authority, maintains the Electronic Data Gathering, Analysis, and Retrieval (EDGAR) system \footnote{\href{https://www.sec.gov/edgar/searchedgar/companysearch}{EDGAR: https://www.sec.gov/edgar/searchedgar/companysearch}}, which is a public database to publicly disclose company fillings. EDGAR is the official data source for Form 10-K. 

Form 10-K document is an annual report required by the SEC to give a comprehensive summary of a company's financial performance. The 10-K is structured under the instruction provided by the SEC. The instruction guides on what to disclose in 4 parts and 22 items of the filing, as shown in Table~\ref{tab:10-K}. 

In addition to the Form 10-K reports, Form 10-Q is a quarterly report required by the SEC to disclose a company's current financial and business conditions for each of the first three quarters in a fiscal year. Compared with Form 10-K, it provides abbreviated information and follows a different structure with 6 prescribed Item sections.

Form 10-K Itemization aims to identify those items and segment the whole document into small pieces for further analysis. The output may not exactly extract 22 items with corresponding document pieces. The reason is that companies may not have anything to disclose about a specific item. This leads to some items missing while following the SEC guidance. Under most circumstances, the Item orders are sequential.

\section{Related Work} \label{sec:relatedWork}
Textual data analysis on financial reports is still an under-explored research topic. Existing algorithms can be roughly categorized into traditional non-deep methods and modern deep learning methods. SPot \cite{zhiqiang_8k} and KGEEF \cite{cheng2020knowledge} are the recent popular deep learning methods.

SPot \cite{zhiqiang_8k} utilizes the NLP technique to extract the operating segments, which shows a significant effect on evaluating the profit and risk of a given company.  However, it only focuses on a few operating segments, which prevents further analysis of other segments with key information. KGEEF \cite{cheng2020knowledge} aims to automatically find the event in the news for quantitative analysis by directly utilizing NLP techniques without domain expert knowledge.

This paper \cite{dyer_evolution_2017} utilizes the Latent Dirichlet Allocation (LDA) to examine specific topics and market trends. \cite{loughran_when_2011} finds a negative word list in a deep model focusing on the 10-K filing returns, trading volume, return volatility, fraud, material weakness, and unexpected earnings. However, there is no comprehensive analysis of the entire reports in those related works. 

Form 10-Q itemization \cite{yancy10q} is a demo itemization system focusing on segmenting the 10-Q report. In comparison, 10-K itemization faces much more challenges than 10-Q reports, as it is the longest among all types of financial reports. Directly applying the Form 10-Q itemization system on 10-K files results in significant performance degradation, since Form 10-K has 4 times more Part numbers and 3 times more Items than Form 10-Q. The total 22 items in 10-K reports cause more detection precision issues. Many more pages and sections mean a lot more candidates to select from, as many as picking 22 right ones from 300 keywords and format-matched samples. The length and amount of Items in Form 10-K reports amplify all the above challenges.

%% file: body/challenge.tex
\section{Challenges} \label{sec:challenge}

In this section, we detail the challenges of designing the itemization algorithm by analyzing why simple solutions fail in segmenting Form 10-K reports. Overall speaking, either rule-based expert knowledge or deep-based models cannot achieve satisfying performance solely. Directly utilizing deep learning methods cannot work well since there are very limited labeled data. It could easily lead to a severe overfitting problem. Directly utilizing expert knowledge cannot work well neither since financial reports have various formats and semantic variations. There could be many unhandled edge cases. More challenges meeting in keyword matching, format matching, and NLP techniques are detailed as follows.

\begin{figure*}[ht]
  \centering
  \shadowsize=1mm
    \color{gray}
    \shadowbox{\fboxsep=0.1mm\fcolorbox{white}{white}{
    \includegraphics[width=0.75\textwidth]
    {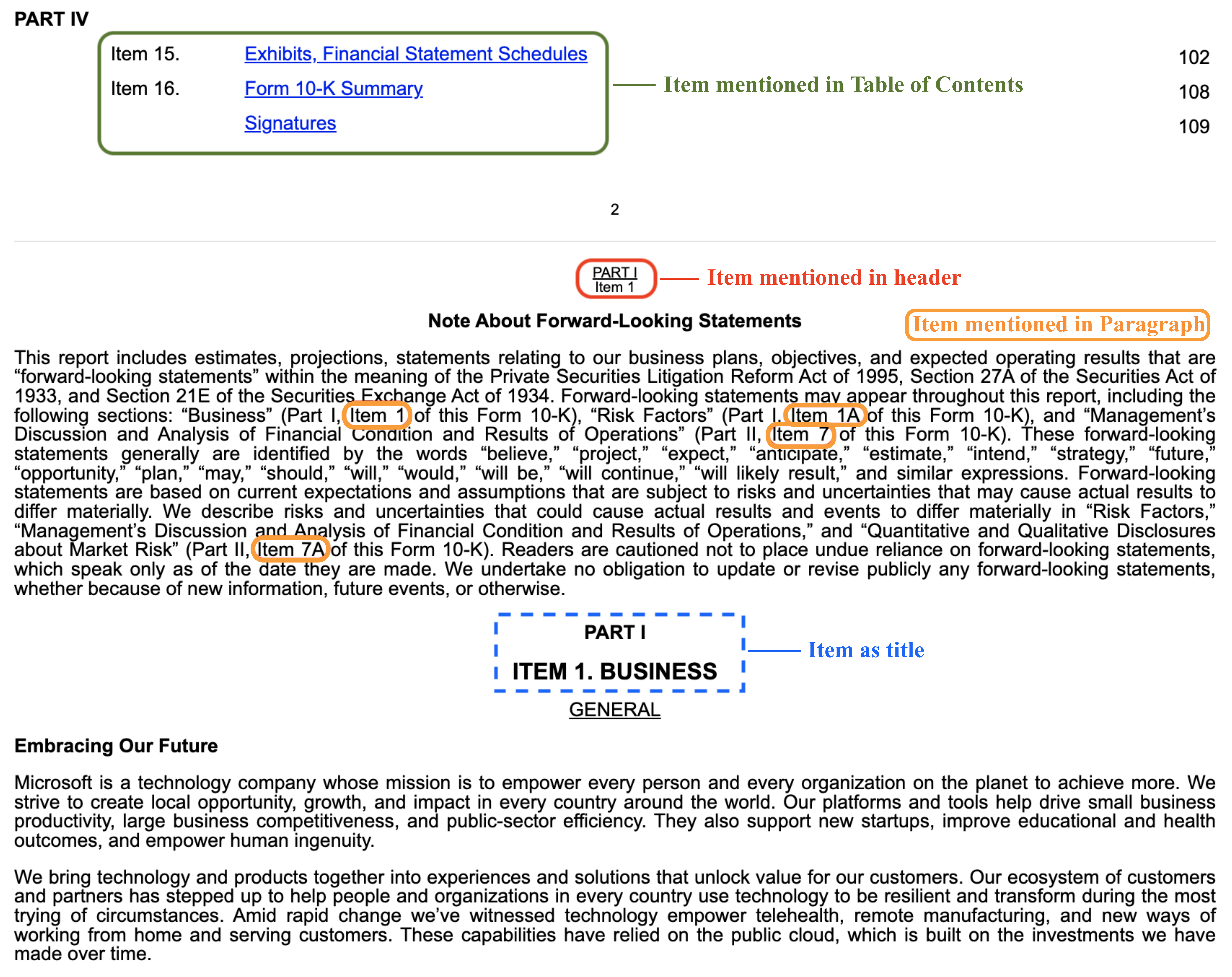}
    }
    }
    \vskip 1em
  \caption{
  A portion of documentation sample from \cite{msft2021}. The blue dot rectangle shows the title of an $\boldsymbol{Item}$, which is the beginning of the target content. The red rectangle shows the $\boldsymbol{Item}$ mentioned as a reference in a paragraph. The green solid rectangle presents a sample of the table of content. 
  }
  \label{fig:example}
\end{figure*}

Keyword matching is the most straightforward baseline to directly match and segment with the Item keyword provided by the regulation of SEC, shown in Table \ref{tab:10-K}. The retrieval rate suffers from a considerable amount of reports with flexibilities in designing and writing each item. Those semantic writing on title name flexibilities can be found in the following perspectives:
\begin{enumerate}
    \item The title for each Item title can be various. It is not necessary to have the exact same name for each section. Only the same semantic meaning is required. Use Item 5 as an example. SEC's official definition is "Item 5. Market for Registrant's Common Equity, Related Stockholder Matters and Issuer Purchases of Equity Securities". It may be presented in different ways by different companies: such as "Global Markets", "Item v", and "Marketing, Distribution and Selected", presented in Figure \ref{fig:example}. These cases cannot be exactly matched with the keyword. A further step with matching part of the keyword may recognize this Item 5. However, there still remain many other unseen cases. 

    \item Some sections are missing. It is not necessary to include every subsection in the financial reports if the company believes nothing needs to disclose or disclose in its own fashion. 
    Take General Electric \cite{ge2021} as an example. The actual “Item 1” is in the section “About General Electric” in page 4-6, “Corporate” is in page 10-15, and “Operating Segments” in page 83-84. Many companies like this may write their own structure and style \cite{ge2021}. The identified semantics structure can be only a substructure of the standard structure. 
\end{enumerate}

Meanwhile, formats in HTML can also provide essential information on recognizing different items. For example, we could design a simple format matching algorithm to match the Item with tab <h> or a larger font size. However, in the real-world case, the format is much more diverse and inconsistent across filings, which cannot easily be handled by simple format rules. 


Moreover, combining keyword plus format may match several hundreds of them, while we only desire 22 Items retrieved. 
Take Ford \cite{ford2021} as an example. “Item” as the keyword would yield 246 matched results. Adding format matching with a large font size or italic could also retrieve hundreds of matched results. 
The combined search leads to a high false positive retrieval rate. The reason that accounts for the high false positive is that an Item title name does not only appear at the beginning point of an Item. Those keywords can also frequently appear in other places like the Table of Contents, header and footer, mentioned in the paragraph, etc., as shown in Figure \ref{fig:example}. It requires human judgment or a more powerful way to handle those cases. 

Deep models meet the above challenges with very limited labeled data available. Besides, directly applying advanced NLP techniques meet additional challenges since those techniques mainly focus on semantic matching while ignoring identifying word differences with similar semantic information. For example, it can be hard for an NLP model to tell the differences between Items as titles and Items mentioned in headers in Figure \ref{fig:example}. It leads to less retrieval rate with too many words with similar semantic information.

%% file: body/methodology.tex
\section{Pipeline Design} \label{sec:method}

\begin{figure*}
  \includegraphics[width=\textwidth]{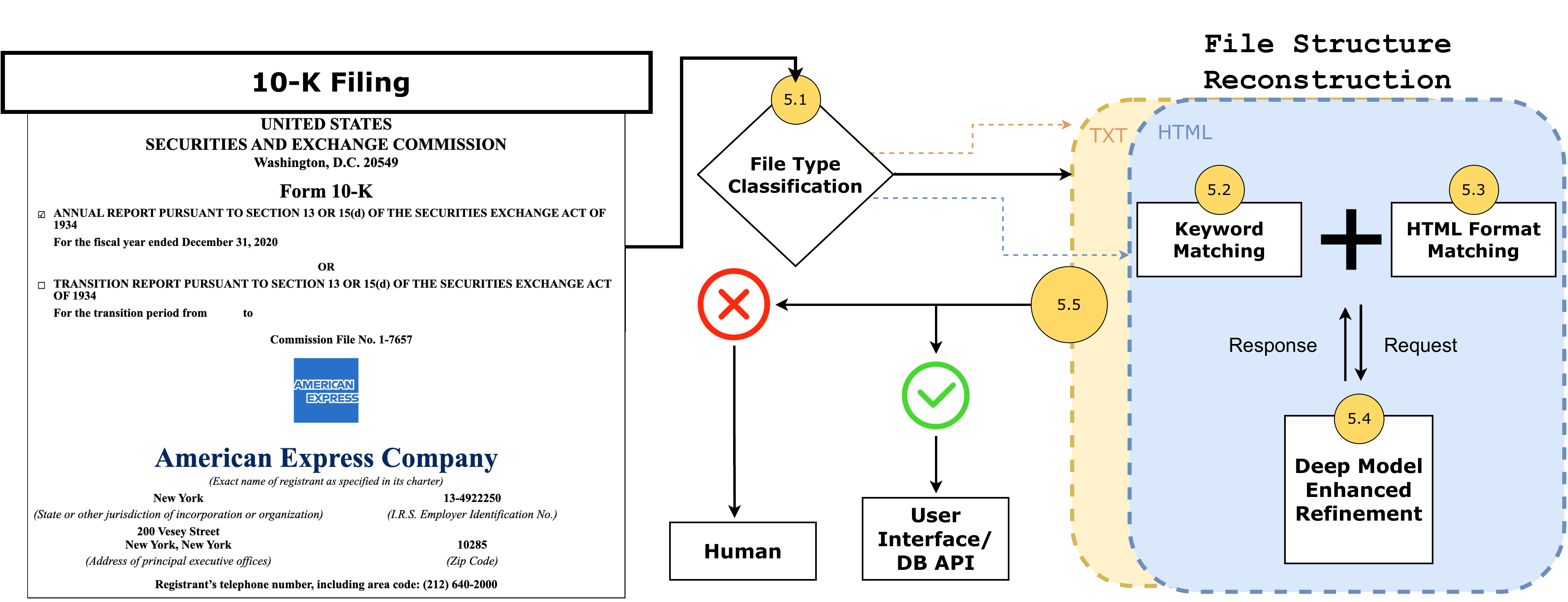}
  \vskip 1em
  \caption{Overview of Form 10-K Itemization system. Components in the dashed rectangular box are tailored to two different data types. The green check mark and red cross indicate whether the process is successful or not. Detailed information about each component can be found in the section denoted in the yellow circles. The cover page of American Express 10-K \cite{amex2021} is used as an example.}
  \label{fig:pipeline}
\end{figure*}

In this section, we first provide an overview of our proposed 10-K itemization system shown in Figure \ref{fig:pipeline}. The pipeline takes the original financial report as the input, extracts the predefined structure, and breaks the whole document into small pieces in a unit of Items as the output. The pipeline consists of six major steps: format detection in \ref{sec:method:format}, keyword matching in \ref{sec:method:keyword}, format matching in \ref{sec:method:html}, deep model enhanced refinement in \ref{sec:method:cv}, file structure reconstruction in \ref{sec:method:structure}, and output segmentation in \ref{sec:method:segmentation}.

Format detection aims to identify whether the financial report is only plain text or HTML at the beginning of the process. It enables us to improve the itemization with the help of type-specific auxiliary information, for example, the tag information in HTML. Different techniques can then be applied to the HTML and plain text separately. 
File structure detection aims to identify the starting and ending position in the document for each Item. It finds a rough scope for where items could lie in. To further refine and identify tons of potential Items, we utilize the deep learning model, to automatically identify and correct for those edge cases instead of using expensive human labor. 
To better utilize the candidate results provided by the above steps, format reconstruction is designed to assemble those results. It leads to a more accurate starting point of items and guarantees the reliability of our itemized data. For those cases without a strong agreement, human experts would be involved in the investigation. Finally, each financial report is segmented using the structure achieved from previous steps, cleaned up, and written to our database for downstream task analysis, which guarantees the cleanness of our itemized data. More details on these steps are provided in the following subsections. 

\subsection{Format Detection} \label{sec:method:format}
Our system is designed to tackle different types and formats of SEC filings. The input of our data pipeline is the “Form 10-K”. The 10-K raw files are obtained from the SEC EDGAR system \footnote{\url{https://www.sec.gov/edgar/searchedgar/companysearch.html}}. These SEC filings are available in two formats - HTML and plain text. Most filings in text format are antiquated, and more recent filings since 2003 are predominantly in HTML format. It is crucial for our system to identify the correct type of format of the input file at the beginning of the process, so the algorithm knows the proper processing procedures to trigger. We formulate this format detection problem as a classification problem with the rule-based classification algorithm. We directly use regular expression matching to search whether the file contains keywords like “html” or “xml”. If keywords are detected, the file is flagged as HTML. Otherwise, it is treated as a plain text file. The segmentation for plain text handling is very straightforward. The Item titles in plain text are usually wrapped around parentheses as an indication of the title, such as “[Item 1]”, while financial reports with HTML format do not have such easily–recognized keywords. The rest of the discussion will focus on the more complicated HTML file format with more complicated formatting and layout options in the following subsections.

\subsection{Keyword Matching} \label{sec:method:keyword}
The target of itemization is to locate the starting and ending positions of each Item in the financial report. As listed in Table~\ref{tab:10-K}, Form 10-K report has at most 22 Items where each Item focuses on a specific financial aspect. Although the SEC specifies the Item numbers and names, there are neither universal Item names nor predetermined segmentation points across Form 10-K reports written by different companies. Revolving on this target, we use a rule-based keyword matching algorithm to locate each Item based on its title.

For Item title pattern matching, we first acquire possible names for each Item by selecting financial reports from 30 different companies. They can provide an initial pattern for further rule design. By traversing more financial reports, we document all possible Item names and build a dictionary. For instance, the names of the predetermined “Item 8: Financial Statements and Supplementary Data” include “Selected Financial Data”, “Financial Statements and Supplementary”, “Non-GAAP Financial Measures”, etc. In accordance with the established keyword dictionary, we use regular expression matching (\{Item number\}.\{Item name\}) to find all the places where the keywords appear. 

\subsection{HTML Format Matching} \label{sec:method:html}
Apart from keyword matching, we propose the format information from HTML to further enhance the matching accuracy. Based on rules summarized by experts in labeled files, we construct another dictionary of special formats that may indicate Item titles. For example, text with (1) a relatively larger size compared with the surrounding, (2) a relatively larger weight compared with the surrounding, (3) centered, (4) italic, (5) length of text in a special format, (6) hyperlink close by is a candidate for special format. 

This format matching reinforces the results from keyword matching. For example, an Item name appearing in its own line with italic type is likely a valid location of the Item starting point. However, the same name in the text may just be a reference to the Item.

\subsection{Deep Model Enhanced Refinement}  \label{sec:method:cv}
The rule-based keyword and format matching procedures can collect almost all the target Items we need in Table \ref{tab:10-K}. However, the false positive rate is too high, which means we extract many irrelevant phrases in different scenarios like (1) table of content, (2) header and footer, (3) reference of the Item in paragraphs, (4) reference of the Item in tables. These are not valid starting points of an Item. In order to fix this problem, we use computer vision algorithms to verify the outputs of previous steps.

The reason why we formulate this step as a computer vision task is two-folded. First, this task can be done with human eyes with little training in financial statements; hence, it is intuitive to use CV to automate it. Second, other non-deep learning-based CV models require input features, while feature engineering is cumbersome and unreliable.

Beginning with image generation, our process imitates screenshots around the items extracted from matching, using “imgkit” library to generate a unified image file with a resolution of 448 x 448. Then, we train image-based binary classifiers, including Resnet50 \cite{resnet}, ViT \cite{dosovitskiy2020image}, and swinTransformer \cite{liu2021swin}, to determine whether given positions are separation points of Items. Use Resnet is because it is a traditional convolutional neural network-based model. It has a solid track record on classification problems. Vision transformers are recent CV advancements. We choose vision transformers for their significantly better at transferring knowledge and generalizing different scenarios, as the example shown in Figure \ref{fig:example}, with little task-specific data. All of the architectures lead to similar resounding performance. Resnet50 has the lowest latency, while Swin Base transformer yields the best generalization performance. 

Finally, the output of the CV classifier can be interpreted as the confidence level or certainty. If the output is above a predefined threshold, the itemized data will be written to our database. Otherwise, the itemized data will be sent to human experts for a second examination and added to the labeled dataset for further system improvement.

\subsection{File Structure Reconstruction} \label{sec:method:structure}
To ensure the robustness of the reconstruction, we combine information from multiple separate detection procedures discussed in \ref{sec:method:keyword}, \ref{sec:method:html}, and \ref{sec:method:cv} on finding the true beginning of an Item.
As illustrated in Figure \ref{fig:pipeline}, each file would pass the above three modules. Keyword matching initially extracts Items based on the predefined Item names and variants. Format matching utilizes the format information to further extract potential Items and validates results from the previous step. After the two matching steps, our system can retrieve 3 up to 20 times more items with exact keyword matched and special HTML formatting than SEC-defined titles, leading to a very high false positive rate. To reduce it, we adopt the CV classifier to confirm whether the extracted Items are correct starting points. 

There may exist some uncertainty for some edge cases in file structure reconstruction. It indicates that the segmentation in the above steps may be incorrect. To ensure the correctness of output data, those data with uncertainty would be sent to human experts for additional verification.

\subsection{Output Segmentation} \label{sec:method:segmentation}
After the above steps, we could find locations of potential Item starting locations in each filing with high confidence. 
Given the starting and ending positions of all Items, noise can still be found in the segmentation result, i.e., header, footer, and page break. 
Moreover, the original format in HTML may also provide an obstacle to build a human-friendly visualization. 
If we directly utilize the original format, there may exist some paragraphs mixed with different formats.  
Facing the above difficulties, we add post-process techniques to reformat the document and remove the noisy words.



%% file: body/result.tex
\section{Experiments} \label{sec:result}
In this section, we implement our pipeline to analyze the real-world Form 10-K reports. We design a series of experiments to answer the following questions. 
\begin{itemize}
    \item Is the Form 10-K itemization system label efficient with the help of expert knowledge and deep model?
    \item How does the Form 10-K itemization system perform in the real-world dataset?
    \item How do different components in the Form 10-K itemization pipeline contribute to the effectiveness?
    \item How does the system perform when numerous requests come simultaneously?  
\end{itemize}

\subsection{Experiment Setting}

We utilize the retrieval rate as the metric to evaluate the performance of our proposed itemization system. 
The document retrieval rate is the proportion of documents whose structures are successfully restored. 
We examine the starting position of each Item restored in this structure. 
The following cases will be identified as the failure: (1) there is an Item written in the report but not in our reconstruction. (2) there is an Item in our construction but not in the original report. (3) any single item’s beginning position is mis-detected. In other words, one incorrectly recognized Item among the 22 items in the Form 10-K report indicates a failure. 
Notice that, the retrieval rate is too strict, which cannot reach high performance even in an ideal case. 
The main reason is that the downstream task tends to conduct analysis on companies in a cross-sectional and chronological fashion. This reconstruction rate provides a lower bound for any Items in this interest.
Meanwhile, this metrics excludes the companies intentionally do not follow the SEC guidance. They have their own content structure, and the content of an Item (defined in the SEC way) may be discussed in different parts of the filing. 
Above all, the document retrieval rate is the proportion of documents whose structures are completely successfully restored.


\subsection{Efficiency on data collection~(Q1)}
Deep Model Enhanced Refinement in Section \ref{sec:method:cv} requires training a deep classifier to identify the most possible starting point of an Item among all candidates given by results feeding from keyword matching and HTML format matching.
An efficient labeling process is employed since there is no available labeled data to train the deep model, 
Details can be found as follows.
We first randomly select filings published between 2005 and 2020. S\&P 500 and Dow 30 composites are assigned a higher sampling weight as they receive more attention from financial analysts and investors. They account for around 75\% of the whole sample.
The selected Form 10-K report is then converted from HTML to an image with a resolution of 448 x 448 utilizing the “imgkit” library. The deep model will then identify whether the Item title is included in the corresponding image.

The labeling process is then implemented, which aims to ensure the diversity of the labeled data with less human effort.
Revolving on this goal, we first label a small subset of 600 filings to train the deep model. Those data serve as the initial data to train the model. The trained model can then be utilized as 
study of distribution on the unlabeled data.



In the first step of labeling, we create a small dataset by randomly generating images of parts of filings. We manually label these images based on whether the image contains the Item title and remove ambiguous images, which results in 300 positive samples and 300 negative samples. We train a ResNet model on this small dataset and apply it to another set of images. We then compare the classification results labeled by the model and ourselves to check for any discrepancies. This comparison allows us to understand what cannot be captured by the initial model due to the limited training data. In particular, we pay special attention to scenarios, discussed in Figure \ref{fig:example}, in the financial reports that are not yet covered in the small dataset, so that we can diversify scenarios to better generate a larger scale dataset.
The deep model can also be utilized to check whether the prediction of the deep model agrees with our labeling process. This comparison allows us to understand what cannot be captured by the initial model due to the limited training data.

In the second step of labeling, with the help of the initial deep model, we label a much larger dataset with more diversified patterns. 
The Item matching algorithm in Sec.\ref{sec:method:keyword} is also utilized to help us filter out a large pool of potential candidates. 
6,685 positive samples are selected from the potential candidates with both human evaluations and the deep model. 
Negative samples are labeled with balancing the occurrence of cases shown in Figure \ref{fig:example} in our dataset similar to real data. 
Notice that, we include many negative samples with the Item keyword, but not appear in the Item title. Those negative samples are not possibly classified correctly with keyword matching. 
We can then a labeled dataset with \mx{6,685} True samples and \mx{4,007} False samples, while 9,567 for training and 1,125 for validation and test, as shown in Table \ref{tab:result:cvlabel}. 
With the above small dataset, we are able to train a deep model with satisfying performance.

\begin{table}[t]
\caption{Distribution of Labeled Data For CV Model}
\label{tab:result:cvlabel}
\begin{tabular}{lll} \hline \hline
      & Training & Testing \\ \hline
TRUE  & 5979  & 706  \\
FALSE & 3588  & 419  \\ \hline \hline
\end{tabular}
\end{table}

\subsection{Effectiveness on Form 10-K Itemization pipeline (Q2)}
We verify the effectiveness of Form 10-K Itemization by checking whether the entire pipeline can identify the structure correctly.
We show how the items are retrieved from both the document and Item level. The results can be found in Table \ref{tab:result:comparison}. 
McDonald and LDA are two baselines which only focuses on extracting on one specific item. We only show their performance on the corresponding item.
10-Q itemization is an itemization system design on the Form 10-Q reports. 
We can see that, our model can achieve an overall item-level retrieval rate of 93\%, while the existing methods cannot perform well. The main reason for why Form 10-Q Itemization cannot perform well is that the Form 10-Q document has much simpler document structure and shorter document than the Form 10-Q reports.





\begin{table}[t]
\caption{The comparison among retrieval rates of proposed system and benchmarks.}
\label{tab:result:comparison}
\begin{tabular}{llll} \hline \hline 
                            & Data Range & Target Item  & Retrieval Rate \\  \hline
McDonald \cite{loughran_when_2011}            & 1994-2008  & Item 1       & 75\%           \\
LDA \cite{dyer_evolution_2017}                & 1996-2013  & Item 7       & 27\%           \\
10-Q Itemization \cite{yancy10q}    & 2005-2020  & All 22 Items & 52\%            \\
10-K Itemization & 2005-2020  & All 22 Items & 93\%          \\ \hline \hline
\end{tabular}
\end{table}

\subsection{Ablation Study~(Q3)}
We conduct an ablation study to further verify the effectiveness of each step in our proposed pipeline. 
Experiment results on item-level True Positive and False Positive are shown in Table \ref{tab:result:ablation}. To further investigate standalone CV model, we experiment all trained models by feeding the data without prior keyword or format matching information, shown in Table \ref{tab:result:cvModelComparison}.

\begin{table}[t]
\caption{Component Analysis: Using CV alone without prior Item keyword or format matching may lead to massive errors on the itemization system. There are only 14 positive Item title, and all models classify way more than true label.}
\label{tab:result:cvModelComparison}
\begin{tabularx}{0.5\textwidth}{l*{6}{c}} \hline \hline
            & Resnet & ViT & ViT  & Swim & Swim\\
            &        & Base & Large & Base & Large\\ \hline
True Positive  & 14  & 10  & 13  & 14 & 14  \\
False Positive & 29  & 18  & 33  & 8  & 17  \\ \hline \hline
\end{tabularx}
\end{table}

\begin{table}[t]
\caption{Performance Analysis on Different Components of the Form 10-K Itemization System.}
\label{tab:result:ablation}
\begin{tabularx}{0.5\textwidth}{l*{4}{c}} \hline \hline 

        & Rule  & CV & Rule + CV \\ 
        \hline 
Correctly Identified Target      & 2093 & 14 & 2093 \\
Incorrectly Identified as Target & 5659  & 8  & 1  \\
Total Positive Label             & 2261 & 14 & 2261 \\
Total Negative Label             & 6659  & 986 & 6659  \\
\hline 
Total Sampled                    & 8752 & 1000 & 8752 \\

\hline \hline
\end{tabularx}
\end{table}

Rule is the pipeline without the deep model structure refinement (Sec.~\ref{sec:method:cv}), and CV refers to the pipeline without keyword matching (Sec.~\ref{sec:method:keyword}) and HTML format matching (Sec.~\ref{sec:method:html}). 

We can find that the Rule method has a high recall but a very low precision. This is because Rule Based Only method performs an exhaustive search of all the occurrences of items in a document but does not have the ability to distinguish a true Item title from Items referred to in the text. Hence, many irrelevant items are retrieved. 
In contrast, CV method has a high precision but a very low recall. 
Although the CV model has the ability to separate true Item titles from irrelevant items based on features embedded in the images, it shows weakness in identifying possible items. 

To further measure the impact, we conduct an experiments on all well trained CV models. All models run on data that does not have prior keyword or format matching information. It means that the input does not guaranteed to have Item section or special format in it. The dataset is created by randomly picking a spot, generating an image and asking CV to classify if the Item title in there. There are total 1000 data points are randomly generated from the same evaluation space. It leads to 14 positive and 986 negative data. Despite CV models having very high accuracy, almost all CV models mistakenly classify double amount of Items title, as shown in \ref{tab:result:cvModelComparison}. 
Using CV alone without prior Item keyword matching thus may lead to massive errors on the itemization system.

Above all, we can see that different components of the baseline provide a complementary effect to the overall pipeline. 
The integration of these components leads to both  high precision and high recall.

\subsection{System Efficiency~(Q4)}
Our system is deployed in a way ready for large scale Machine Learning and financial data analytics. The system could handle large amount of concurrent requests with minimum latency. 

Specifically, our deployment could handle more than 10,000 queries per second (QPS) tested by Apache Bench. The experiment is conducted in a cluster outside of our system deployment cluster, as discussed in Section\ref{sec:demo}, to simulate the request sending from our clients. We are able to achieve mean QPS of 12,800.

In order to adapt near real time financial analysis, our systems is able to provide data with low latency. The 90\% data (P90) responses within 52 milliseconds. 

%% file: body/demo.tex
\section{User Interface and Systems Deployment} \label{sec:demo}

In this section, we discuss the user interface and the back-end system deployment.
The user interface is to demonstrate our proposed itemization system, and we describe the high-level design of our high availability back-end deployment.

Figures \ref{fig:demo} and \ref{fig:demo:reader} demonstrate the user interface of our Form 10-K itemization system, where researchers and financial analysts can easily access the system and segment financial reports. 

The landing page (Figure \ref{fig:demo}) provides a few sample filings and options to analyze any filings users are interested in.
They can achieve that by providing the link from the SEC EDGAR website. After submitting the itemization request, users will be directed to a reader page (Figure \ref{fig:demo:reader}). All the items retrieved by our Form 10-K Itemization system are displayed on the right. Users can click the Item number to view its content on the left. By clicking the ``Export'' buttons on the top, users can save the results immediately. The ``Open New Form'' button will redirect users to the landing page.

\begin{figure}[t]
  \centering
  \shadowsize=1mm
    \color{gray}
    \shadowbox{\fboxsep=0.1mm\fcolorbox{white}{white}{
    \includegraphics[width=0.6\linewidth]{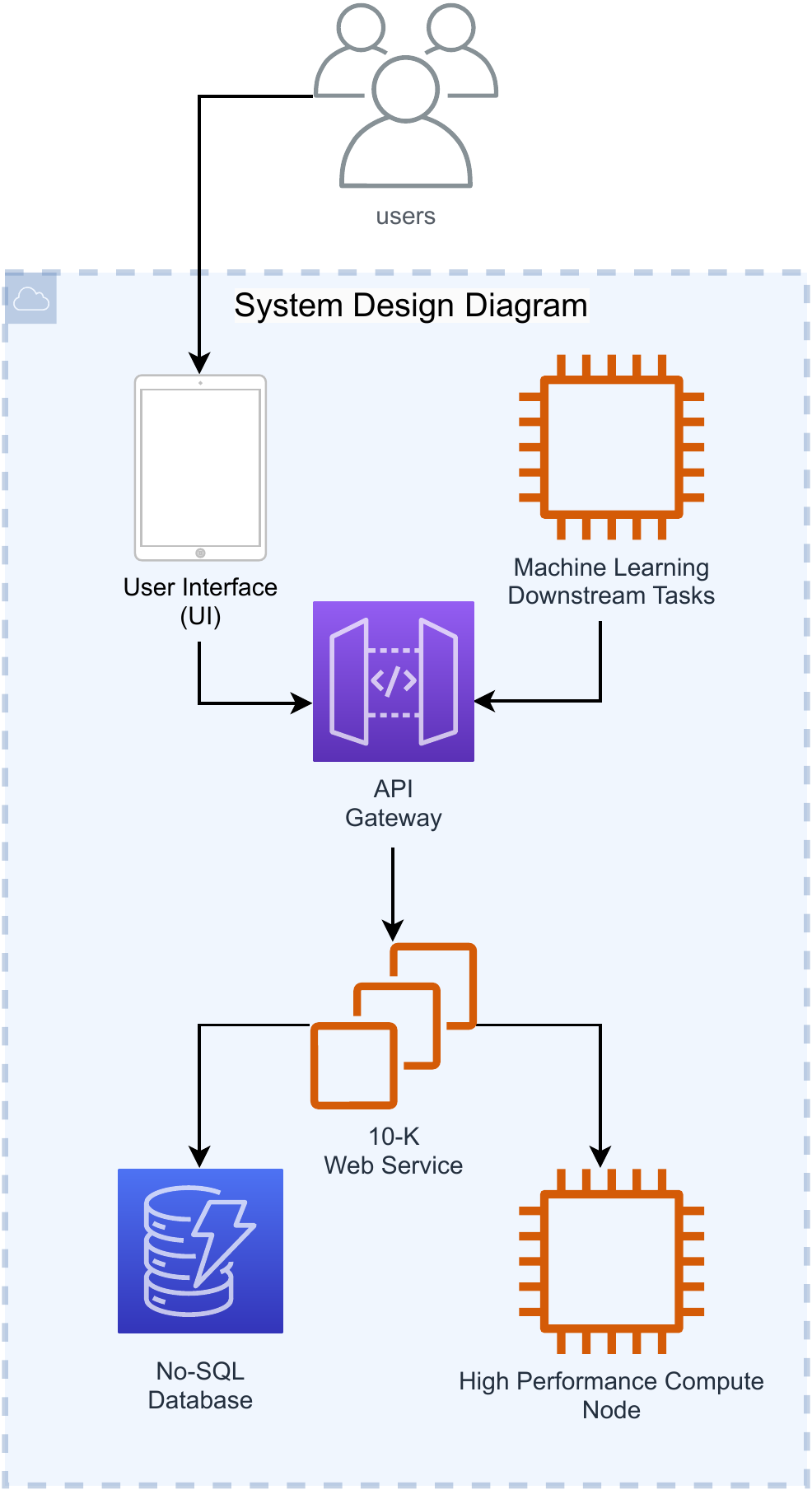}}
    }
  \caption{
    High-level design of 10-K Itemization back end system.
  }
  \label{fig:system-design}
\end{figure}

Figure \ref{fig:system-design} shows the high-level design of our high availability and low latency back-end systems. There are two major use cases. First, the requests are coming from User Interface. The number of requests and throughput are generally very low because it designs to operate by humans. Second, the requests are coming from 1) real-time financial analytics or decision making 2) machine learning training and inference. The requests coming from these two use cases are tremendous and require low response latency in some cases. For this reason, our system is designed to handle 10,000 requests per second while having a P90 latency of 52 ms. 

In detail, the requests are first landing at our API gateway. Then it will be load balanced into multiple web servers hosting our 10-K service. 
The 10-K service will first check if the filing has been processed before. If so, the No-SQL database storing the record of items in the form of key-value pair would directly respond to the client. This avoids processing the same financial filing if we have a record to respond immediately. 
If the financial fining has never been processed before, it will be forwarded from the 10-K web service to our high-performance computing instance. The instance has the strongest CPU and GPU in our cluster hosting our proposed itemization system. Finally, the itemized results would be responded to the client.

%% file: body/conclusion.tex
\section{Conclusion \& future work} \label{sec:conclusion}

In this research, we present a novel 10-K Itemization system with the aim of automating the segmentation of annual reports of US companies. It is the first-of-its-kind itemization system on Form 10-K. This system boasts full coverage of the Item sections in 10-K reports with a remarkable 93\% retrieval rate. Additionally, the system is economically efficient as it requires a limited number of data samples labeled by finance experts. This provides the system extra reliability and accuracy in addition to the automatic system. Furthermore, the system is easily generalized to all types of financial reports beyond Form 10-K. The system has been successfully deployed, with remarkable availability and the capability of serving over 10,000 simultaneous customer requests.

The focus of the itemization system is Form 10-K report, and the proposed system can be generalized to other financial reports, of which the SEC gives guidance on the structure. However, there are other documents that do not have a pre-defined structure but are also important to investors and researchers. How to generalize the current system to process those documents is an interesting topic for future work. We think this generalization will meet technical challenges. For example, the current system relies on prior knowledge of Item titles. However, it may be hard to incorporate all this information if we would like to design a comprehensive system. It is a question of future research to investigate whether we can itemize the documents without prior knowledge. In addition, the current system allows us to retrieve high-quality 10-K data, which enables many downstream NLP tasks. For example, a chronological comparison of the same Item of the same company or a cross-sectional comparison of the same Item of different companies can be done to analyze the changes in the company management.     

%% file: main.bbl

\begin{thebibliography}{22}


\ifx \showCODEN    \undefined \def \showCODEN     #1{\unskip}     \fi
\ifx \showDOI      \undefined \def \showDOI       #1{#1}\fi
\ifx \showISBNx    \undefined \def \showISBNx     #1{\unskip}     \fi
\ifx \showISBNxiii \undefined \def \showISBNxiii  #1{\unskip}     \fi
\ifx \showISSN     \undefined \def \showISSN      #1{\unskip}     \fi
\ifx \showLCCN     \undefined \def \showLCCN      #1{\unskip}     \fi
\ifx \shownote     \undefined \def \shownote      #1{#1}          \fi
\ifx \showarticletitle \undefined \def \showarticletitle #1{#1}   \fi
\ifx \showURL      \undefined \def \showURL       {\relax}        \fi
\providecommand\bibfield[2]{#2}
\providecommand\bibinfo[2]{#2}
\providecommand\natexlab[1]{#1}
\providecommand\showeprint[2][]{arXiv:#2}

\bibitem[Cheng et~al\mbox{.}(2020)]%
        {cheng2020knowledge}
\bibfield{author}{\bibinfo{person}{Dawei Cheng}, \bibinfo{person}{Fangzhou
  Yang}, \bibinfo{person}{Xiaoyang Wang}, \bibinfo{person}{Ying Zhang}, {and}
  \bibinfo{person}{Liqing Zhang}.} \bibinfo{year}{2020}\natexlab{}.
\newblock \showarticletitle{Knowledge graph-based event embedding framework for
  financial quantitative investments}. In \bibinfo{booktitle}{\emph{Proceedings
  of the 43rd International ACM SIGIR Conference on Research and Development in
  Information Retrieval}}. \bibinfo{pages}{2221--2230}.
\newblock


\bibitem[Company(2021a)]%
        {amex2021}
\bibfield{author}{\bibinfo{person}{American~Express Company}.}
  \bibinfo{year}{2021}\natexlab{a}.
\newblock \bibinfo{booktitle}{\emph{axp-20201231}}.
\newblock
\urldef\tempurl%
\url{https://www.sec.gov/ix?doc=/Archives/edgar/data/4962/000000496221000013/axp-20201231.htm}
\showURL{%
Retrieved Feb 2, 2023 from \tempurl}


\bibitem[Company(2021b)]%
        {ford2021}
\bibfield{author}{\bibinfo{person}{Ford~Motor Company}.}
  \bibinfo{year}{2021}\natexlab{b}.
\newblock \bibinfo{booktitle}{\emph{f1231201910-k}}.
\newblock
\urldef\tempurl%
\url{https://www.sec.gov/ix?doc=/Archives/edgar/data/37996/000003799620000010/f1231201910-k.htm}
\showURL{%
Retrieved Feb 2, 2023 from \tempurl}


\bibitem[Corporation(2020)]%
        {mcd2020}
\bibfield{author}{\bibinfo{person}{McDonald Corporation}.}
  \bibinfo{year}{2020}\natexlab{}.
\newblock \bibinfo{booktitle}{\emph{mcd-20201231}}.
\newblock
\urldef\tempurl%
\url{https://www.sec.gov/ix?doc=/Archives/edgar/data/0000063908/000006390821000013/mcd-20201231.htm}
\showURL{%
Retrieved Feb 2, 2023 from \tempurl}


\bibitem[Dosovitskiy et~al\mbox{.}(2020)]%
        {dosovitskiy2020image}
\bibfield{author}{\bibinfo{person}{Alexey Dosovitskiy}, \bibinfo{person}{Lucas
  Beyer}, \bibinfo{person}{Alexander Kolesnikov}, \bibinfo{person}{Dirk
  Weissenborn}, \bibinfo{person}{Xiaohua Zhai}, \bibinfo{person}{Thomas
  Unterthiner}, \bibinfo{person}{Mostafa Dehghani}, \bibinfo{person}{Matthias
  Minderer}, \bibinfo{person}{Georg Heigold}, \bibinfo{person}{Sylvain Gelly},
  {et~al\mbox{.}}} \bibinfo{year}{2020}\natexlab{}.
\newblock \showarticletitle{An image is worth 16x16 words: Transformers for
  image recognition at scale}.
\newblock \bibinfo{journal}{\emph{arXiv preprint arXiv:2010.11929}}
  (\bibinfo{year}{2020}).
\newblock


\bibitem[Dyer et~al\mbox{.}(2017)]%
        {dyer_evolution_2017}
\bibfield{author}{\bibinfo{person}{Travis Dyer}, \bibinfo{person}{Mark Lang},
  {and} \bibinfo{person}{Lorien Stice-Lawrence}.}
  \bibinfo{year}{2017}\natexlab{}.
\newblock \showarticletitle{The evolution of 10-{K} textual disclosure:
  {Evidence} from {Latent} {Dirichlet} {Allocation}}.
\newblock \bibinfo{journal}{\emph{Journal of Accounting and Economics}}
  \bibinfo{volume}{64}, \bibinfo{number}{2-3} (\bibinfo{year}{2017}),
  \bibinfo{pages}{221--245}.
\newblock
\newblock
\shownote{ISBN: 0165-4101, Elsevier}.


\bibitem[Fama and French(1993)]%
        {fama1993common}
\bibfield{author}{\bibinfo{person}{Eugene~F Fama} {and}
  \bibinfo{person}{Kenneth~R French}.} \bibinfo{year}{1993}\natexlab{}.
\newblock \showarticletitle{Common risk factors in the returns on stocks and
  bonds}.
\newblock \bibinfo{journal}{\emph{Journal of financial economics}}
  \bibinfo{volume}{33}, \bibinfo{number}{1} (\bibinfo{year}{1993}),
  \bibinfo{pages}{3--56}.
\newblock


\bibitem[Fama and French(2015)]%
        {fama2015five}
\bibfield{author}{\bibinfo{person}{Eugene~F Fama} {and}
  \bibinfo{person}{Kenneth~R French}.} \bibinfo{year}{2015}\natexlab{}.
\newblock \showarticletitle{A five-factor asset pricing model}.
\newblock \bibinfo{journal}{\emph{Journal of financial economics}}
  \bibinfo{volume}{116}, \bibinfo{number}{1} (\bibinfo{year}{2015}),
  \bibinfo{pages}{1--22}.
\newblock


\bibitem[Gu et~al\mbox{.}(2020)]%
        {gu2020empirical}
\bibfield{author}{\bibinfo{person}{Shihao Gu}, \bibinfo{person}{Bryan Kelly},
  {and} \bibinfo{person}{Dacheng Xiu}.} \bibinfo{year}{2020}\natexlab{}.
\newblock \showarticletitle{Empirical asset pricing via machine learning}.
\newblock \bibinfo{journal}{\emph{The Review of Financial Studies}}
  \bibinfo{volume}{33}, \bibinfo{number}{5} (\bibinfo{year}{2020}),
  \bibinfo{pages}{2223--2273}.
\newblock


\bibitem[He et~al\mbox{.}(2016)]%
        {resnet}
\bibfield{author}{\bibinfo{person}{Kaiming He}, \bibinfo{person}{Xiangyu
  Zhang}, \bibinfo{person}{Shaoqing Ren}, {and} \bibinfo{person}{Jian Sun}.}
  \bibinfo{year}{2016}\natexlab{}.
\newblock \showarticletitle{Deep Residual Learning for Image Recognition}.
  \bibinfo{pages}{770--778}.
\newblock
\urldef\tempurl%
\url{https://doi.org/10.1109/CVPR.2016.90}
\showDOI{\tempurl}


\bibitem[Huang and Li(2011)]%
        {huang2011multilabel}
\bibfield{author}{\bibinfo{person}{Ke-Wei Huang} {and} \bibinfo{person}{Zhuolun
  Li}.} \bibinfo{year}{2011}\natexlab{}.
\newblock \showarticletitle{A multilabel text classification algorithm for
  labeling risk factors in SEC form 10-K}.
\newblock \bibinfo{journal}{\emph{ACM Transactions on Management Information
  Systems (TMIS)}} \bibinfo{volume}{2}, \bibinfo{number}{3}
  (\bibinfo{year}{2011}), \bibinfo{pages}{1--19}.
\newblock


\bibitem[Inc.(2021a)]%
        {ge2021}
\bibfield{author}{\bibinfo{person}{General~Electric Inc.}}
  \bibinfo{year}{2021}\natexlab{a}.
\newblock \bibinfo{booktitle}{\emph{ge-20211231}}.
\newblock
\urldef\tempurl%
\url{https://www.sec.gov/ix?doc=/Archives/edgar/data/40545/000004054522000008/ge-20211231.htm}
\showURL{%
Retrieved Feb 2, 2023 from \tempurl}


\bibitem[Inc.(2021b)]%
        {msft2021}
\bibfield{author}{\bibinfo{person}{Microsoft Inc.}}
  \bibinfo{year}{2021}\natexlab{b}.
\newblock \bibinfo{booktitle}{\emph{msft-10k\_20210630}}.
\newblock
\urldef\tempurl%
\url{https://www.sec.gov/ix?doc=/Archives/edgar/data/789019/000156459021039151/msft-10k_20210630.htm}
\showURL{%
Retrieved Feb 2, 2023 from \tempurl}


\bibitem[Li et~al\mbox{.}(2013)]%
        {li2013measure}
\bibfield{author}{\bibinfo{person}{Feng Li}, \bibinfo{person}{Russell
  Lundholm}, {and} \bibinfo{person}{Michael Minnis}.}
  \bibinfo{year}{2013}\natexlab{}.
\newblock \showarticletitle{A measure of competition based on 10-K filings}.
\newblock \bibinfo{journal}{\emph{Journal of Accounting Research}}
  \bibinfo{volume}{51}, \bibinfo{number}{2} (\bibinfo{year}{2013}),
  \bibinfo{pages}{399--436}.
\newblock


\bibitem[Liu et~al\mbox{.}(2023)]%
        {liutracking}
\bibfield{author}{\bibinfo{person}{Rong Liu}, \bibinfo{person}{Jujun Huang},
  {and} \bibinfo{person}{Zhongju Zhang}.} \bibinfo{year}{2023}\natexlab{}.
\newblock \showarticletitle{Tracking disclosure change trajectories for
  financial fraud detection}.
\newblock \bibinfo{journal}{\emph{Production and Operations Management}}
  \bibinfo{volume}{32}, \bibinfo{number}{2} (\bibinfo{year}{2023}),
  \bibinfo{pages}{584--602}.
\newblock


\bibitem[Liu et~al\mbox{.}(2021)]%
        {liu2021swin}
\bibfield{author}{\bibinfo{person}{Ze Liu}, \bibinfo{person}{Yutong Lin},
  \bibinfo{person}{Yue Cao}, \bibinfo{person}{Han Hu}, \bibinfo{person}{Yixuan
  Wei}, \bibinfo{person}{Zheng Zhang}, \bibinfo{person}{Stephen Lin}, {and}
  \bibinfo{person}{Baining Guo}.} \bibinfo{year}{2021}\natexlab{}.
\newblock \showarticletitle{Swin transformer: Hierarchical vision transformer
  using shifted windows}. In \bibinfo{booktitle}{\emph{Proceedings of the
  IEEE/CVF international conference on computer vision}}.
  \bibinfo{pages}{10012--10022}.
\newblock


\bibitem[Loughran and McDonald(2011)]%
        {loughran_when_2011}
\bibfield{author}{\bibinfo{person}{Tim Loughran} {and} \bibinfo{person}{Bill
  McDonald}.} \bibinfo{year}{2011}\natexlab{}.
\newblock \showarticletitle{When is a liability not a liability? {Textual}
  analysis, dictionaries, and 10‐{Ks}}.
\newblock \bibinfo{journal}{\emph{The Journal of Finance}}
  \bibinfo{volume}{66}, \bibinfo{number}{1} (\bibinfo{year}{2011}),
  \bibinfo{pages}{35--65}.
\newblock
\newblock
\shownote{ISBN: 0022-1082 Publisher: Wiley Online Library}.


\bibitem[Ma et~al\mbox{.}(2020)]%
        {zhiqiang_8k}
\bibfield{author}{\bibinfo{person}{Zhiqiang Ma}, \bibinfo{person}{Steven
  Pomerville}, \bibinfo{person}{Mingyang Di}, {and} \bibinfo{person}{Armineh
  Nourbakhsh}.} \bibinfo{year}{2020}\natexlab{}.
\newblock \showarticletitle{SPot: A Tool for Identifying Operating Segments in
  Financial Tables}. In \bibinfo{booktitle}{\emph{Proceedings of the 43rd
  International ACM SIGIR Conference on Research and Development in Information
  Retrieval}} (Virtual Event, China) \emph{(\bibinfo{series}{SIGIR '20})}.
  \bibinfo{publisher}{Association for Computing Machinery},
  \bibinfo{address}{New York, NY, USA}, \bibinfo{pages}{2157–2160}.
\newblock
\showISBNx{9781450380164}
\urldef\tempurl%
\url{https://doi.org/10.1145/3397271.3401406}
\showDOI{\tempurl}


\bibitem[Mushtaq et~al\mbox{.}(2022)]%
        {mushtaq2022financial}
\bibfield{author}{\bibinfo{person}{Rizwan Mushtaq}, \bibinfo{person}{Ammar~Ali
  Gull}, \bibinfo{person}{Yasir Shahab}, {and} \bibinfo{person}{Imen
  Derouiche}.} \bibinfo{year}{2022}\natexlab{}.
\newblock \showarticletitle{Do financial performance indicators predict 10-K
  text sentiments? An application of artificial intelligence}.
\newblock \bibinfo{journal}{\emph{Research in International Business and
  Finance}}  \bibinfo{volume}{61} (\bibinfo{year}{2022}),
  \bibinfo{pages}{101679}.
\newblock


\bibitem[SEC(2023)]%
        {tenk-guideline}
\bibfield{author}{\bibinfo{person}{SEC}.} \bibinfo{year}{2023}\natexlab{}.
\newblock \bibinfo{booktitle}{\emph{form10-k}}.
\newblock
\urldef\tempurl%
\url{https://www.sec.gov/files/form10-k.pdf}
\showURL{%
Retrieved Feb 2, 2023 from \tempurl}


\bibitem[Yang et~al\mbox{.}(2018)]%
        {yang2018corporate}
\bibfield{author}{\bibinfo{person}{Rong Yang}, \bibinfo{person}{Yang Yu},
  \bibinfo{person}{Manlu Liu}, {and} \bibinfo{person}{Kean Wu}.}
  \bibinfo{year}{2018}\natexlab{}.
\newblock \showarticletitle{Corporate risk disclosure and audit fee: A text
  mining approach}.
\newblock \bibinfo{journal}{\emph{European Accounting Review}}
  \bibinfo{volume}{27}, \bibinfo{number}{3} (\bibinfo{year}{2018}),
  \bibinfo{pages}{583--594}.
\newblock


\bibitem[Zhang et~al\mbox{.}(2021)]%
        {yancy10q}
\bibfield{author}{\bibinfo{person}{Yanci Zhang}, \bibinfo{person}{Tianming Du},
  \bibinfo{person}{Yujie Sun}, \bibinfo{person}{Lawrence Donohue}, {and}
  \bibinfo{person}{Rui Dai}.} \bibinfo{year}{2021}\natexlab{}.
\newblock \showarticletitle{Form 10-Q Itemization}. In
  \bibinfo{booktitle}{\emph{Proceedings of the 30th ACM International
  Conference on Information \& Knowledge Management}} (Virtual Event,
  Queensland, Australia) \emph{(\bibinfo{series}{CIKM '21})}.
  \bibinfo{publisher}{Association for Computing Machinery},
  \bibinfo{address}{New York, NY, USA}, \bibinfo{pages}{4817–4822}.
\newblock
\showISBNx{9781450384469}
\urldef\tempurl%
\url{https://doi.org/10.1145/3459637.3481989}
\showDOI{\tempurl}


\end{thebibliography}
